\documentclass[pre,superscriptaddress]{revtex4}

\usepackage{graphicx,t1enc}
\usepackage{epsfig}
\usepackage{epstopdf}
\usepackage{amssymb,amsmath}
\usepackage{xcolor}
\usepackage{amstext}
\usepackage{amssymb}
\usepackage{graphicx} 
\usepackage{esint}
\usepackage{ulem}
\usepackage{subcaption}
\usepackage{placeins}
\usepackage{hyperref}

\definecolor{blue}{rgb}{0,0,1}

\begin{document}

\title[]{Analysis of non pharmaceutical interventions with SIR epidemic models: decreasing the infection peak vs. 
minimizing the epidemic size.}

\author{Eric Roz\'an}
\affiliation{Instituto Balseiro, Universidad Nacional de Cuyo y Comisión Nacional de Energía Atómica, Bariloche, Argentina.}
\affiliation{Consejo Nacional de Investigaciones Cient\'{\i}ficas y T\'ecnicas (CONICET), Argentina.} 

\author{Marcelo Kuperman}
\affiliation{Instituto Balseiro, Universidad Nacional de Cuyo y Comisión Nacional de Energía Atómica, Bariloche, Argentina.}
\affiliation{Consejo Nacional de Investigaciones Cient\'{\i}ficas y T\'ecnicas (CONICET), Argentina.} 
\affiliation{Centro At\'omico Bariloche (CNEA), (8400) Bariloche, R\'{\i}o Negro, Argentina.}

\author{Sebasti\'an Bouzat}\thanks{email: sebabouzat@gmail.com}
\affiliation{Consejo Nacional de Investigaciones Cient\'{\i}ficas y T\'ecnicas (CONICET), Argentina.} 
\affiliation{Centro At\'omico Bariloche (CNEA), (8400) Bariloche, R\'{\i}o Negro, Argentina.}
\begin{abstract}
This study investigates the influence of different types of  non-pharmaceutical interventions (NPIs) on epidemic progression using SIR compartmental models. We analyze the optimization of two distinct targets: the final epidemic size and the infection peak, particularly how they respond to variations
in the initiation time of the NPIs. We derive analytical approximations for the critical points of the infection curve of the standard mean-field SIR model with NPIs, and for the epidemic size, enabling a systematic comparison. The analytical results reveal the existence of six different allowed scenarios for the evolution
of the epidemic with a single NPI. Furthermore, by employing degree-based mean-field network models, we distinguish
between NPIs that decrease the transmission rate (individual and environmental measures) and those
that reduce social contacts (lock down measures). We find that, when assuming equal effects on the reproductive number, the former are more efficient in reducing the final epidemic size. Meanwhile, the effectivities of both types of NPIs differ in reducing primary and secondary peaks. The results for all models consistently confirm that minimizing the infection peak requires earlier implementation of the NPI than minimizing the epidemic size, offering new insights for strategic public health timing.

\end{abstract}
\maketitle

\section{Introduction}
Mathematical models have played a significant role for many decades in understanding the complex mechanisms involved in the spread of infectious diseases~\cite{kerm1,1001,sartwell,Hethcote,keeling}. In particular, the application of statistical physics and network theory has significantly advanced our comprehension of how the underlying social structure influences the epidemiological transition \cite{Andersson,pastorsatorras}. Since 2020, the COVID-19 pandemic has triggered an unprecedented surge in the production of studies focused on epidemic modeling (see for example \cite{vaccination,perra,RKBincub} and the references therein). Undoubtedly, theoretical and computational modeling tools have become essential instruments for evaluating the impact of public health policies and the effectiveness of diverse intervention strategies, including, for instance, vaccination \cite{vaccination} and Non Pharmaceutical Interventions (NPIs) \cite{yan,perra}.

One of the most relevant modeling frameworks is that of compartmental models, such as SIR models~\cite{kerm1}, for which the populations are separated in compartments corresponding to Susceptible, Infected and Recovered individuals. Many different versions of compartmental models have been developed, including additional compartments for exposed, vaccinated, isolated and also separation by ages, among other examples~\cite{1001}.    

While quarantine practices date back to biblical times and have been employed throughout history, the modern implementation of Non-Pharmaceutical Interventions (NPIs) presents a significant challenge, due to the inherent trade-off between epidemiological benefits and their profound socio-economic and psychological costs. Although these measures, such as quarantines, are often necessary to curb pathogen transmission, they inevitably entail severe disruptions to economic stability and to the mental well-being of the population. This complexity demands a deep understanding of how such interventions operate and how they can be optimized according to the specific needs of a society and the biological particularities of the disease. 

Recently, Atias and Assaf \cite{Atias2025} have addressed the question of finding the optimal initiation time for a single NPI to minimize the final outbreak size of epidemics. For this, they have taken a fundamental perspective by considering a basic compartmental SIR model and examining the cases of well-mixed populations and heterogeneous networks. Previously, most of the modeling studies on quarantines and NPIs analyzed the influence of the duration of the interventions and other relevant features \cite{yan}, including effects related to the social structure \cite{Rozan}, the role of alternating \cite{meidan,cornes} and successive \cite{hindes} interventions, to cite a few, without particularly focusing on the importance of tuning the initiation time.

In this work, we extend and complement in two different ways the analysis provided in \cite{Atias2025} concerning the adjustment of the initial time of an NPI. First, we vary the optimization target, as we analyze the problem of minimizing the peak of the infection in addition to that of reducing the epidemic size, and compare the results for the two cases. Second, we consider two different types of NPIs. On the one hand, individual protective and environmental measures that do not directly affect the structure of social contacts, such as hand washing, the use of face masks, or increased ventilation. On the other hand, we analyze NPIs that involve changes in the network of contacts, such as social distancing and stay-at-home orders. While the standard SIR model with well-mixed populations cannot distinguish between these two types of NPIs, degree-based mean-field models such as the one proposed in
\cite{Moreno1} allow us to do it~\cite{Rozan}.

Regarding the choice of the optimization target, it is worth noting that this is not universal and depends heavily on the epidemiological characteristics of the particular disease and the available medical resources. For example, in diseases with high morbidity but relatively low mortality, the primary concern may not be the total number of infected individuals, but rather preventing the saturation of the healthcare system. In such scenarios, minimizing the peak of the infection curve becomes more important. In contrast, if the disease produces high mortality but medical resources are sufficient to treat all patients, the focus should be on minimizing the final outbreak size. 

When dealing with the problem of reducing the infection peak, we developed analytical approximations for the standard SIR model with NPIs that extend the results given in \cite{Atias2025} and may be of relevance for future research in different related subjects.

The organization of the paper is as follows. In section II we present the standard SIR model with a temporary NPI following the steps in \cite{Atias2025}, in section III we study the infection peaks and the critical points, in section IV we develop the analytical approximation that enable us to express the critical points as functions of the system parameters. In section V we present the results for optimization of the infection peaks and compare them with those for minimizing the outbreak size. In section VI we introduce the network model and use it to compare results for the two types of NPIs indicated above. Section VII is devoted to our conclusions and some final remarks.

\section{SIR Model with temporary NPI.} 

In this section we present the SIR model with an NPI of duration $\Delta t$ that we use as a basis for our studies. For this, we follow the steps in \cite{Atias2025}. We begin by considering the standard SIR model~\cite{kerm1}:
\begin{eqnarray}
\label{SIReqs}
\dot{S}&=&-\alpha S I \nonumber \\
\dot{I}&=& \alpha S I - \gamma I \nonumber \\
\dot{R}&=& \gamma I. 
\end{eqnarray}
Here, $S, I,$ and $R$ are the fractions of susceptible, infected and recovered agents, respectively. The parameter $\alpha$ is the contagion rate and $\gamma$ is the recovery rate. The initial condition considered is $I(0)=I_0\ll1, S(0)=1-I_0, R(0)=0$ so that $S(t)+I(t)+R(t)=1$ holds at any time. For this model, the reproduction number is well known to be $\mathcal R_0= \alpha/\gamma$~\cite{kerm1}. 

In \cite{Atias2025}, in order to include an NPI between $t_0>0$ and $t_0+\Delta t$, the authors considered that the contagious rate decreases by a factor $\xi$ $(0<\xi<1)$ during such time interval. Following such assumption, rescaling time $t \to \gamma t$ and defining 
$\beta=\alpha/\gamma$ we get 
\begin{eqnarray}
\label{SIRconxi}
\dot{S}&=&-{\cal B}(t) S I \nonumber \\
\dot{I}&=& {\cal B}(t) S I - I \nonumber \\
\dot{R}&=& I,
\end{eqnarray}
where
\begin{equation}
{\cal B}(t)= \begin{cases} \beta &{\rm for\,\,} t<t_0 \,\,{\rm or}\,\, t>t_0+\Delta t, \\ \beta \, \xi \,\, &{\rm for}\,\, t_0<t<t_0+\Delta t. \end{cases} 
\end{equation}
The constant $\beta=\alpha/\gamma$ coincides with the reproduction number $\mathcal R_0$, which characterizes the propagation properties at the beginning of the epidemics. 
Following again \cite{Atias2025}, we define $S_b=S(t_0)$ as the fraction of susceptible agents at the beginning of the NPI. Note that since the fraction of susceptible individuals -- or rather, its complement, the cummulative cases-- is in practice a measurable quantity, within the model $S_b$ can work as a threshold value that indicates when to apply the NPI. The relation $S(t_0)=S_b$ determines the starting time $t_0$. Thus, we consider $S_b$ instead of $t_0$ as the relevant parameter. The system parameters are $I_0, \beta, \xi, S_b,$ and $ \Delta t$. This work, as \cite{Atias2025}, deals with the problem of finding optimal values of $S_b$ for different assumptions and situations. For simplicity, we define $t_\Delta\equiv t_0+\Delta t$ to represent the time at which the NPI ends.

In order to find analytical solutions for Eqs. (\ref{SIRconxi}), as done in \cite{Atias2025} and \cite{julicher2020} we consider the scaled time $\tau$ defined such that $\dot{\tau}={\cal B} I$. This leads to the simple expression $S(\tau)=(1-I_0) \exp(-\tau)$ for $S(\tau)$ \cite{Atias2025}. The solutions for $I(\tau)$ and $R(\tau)$ are more intricate \cite{Atias2025} than that for $S(\tau)$, since they have to be defined in different regions of $\tau$. Moreover, they depend on the duration of the NPI ($\Delta t$) measured in the time scale $\tau$, referred to as $\Delta \tau$ \cite{Atias2025}, for which there is no explicit expression. The solutions can be written as \cite{Atias2025} 

\begin{eqnarray}
\label{IRdetau}
I(\tau)&=&1-(1-I_0) e^{-\tau} - R(\tau), \nonumber \\
R(\tau)&=& \begin{cases} \frac{\tau}{\beta} & \tau \le \tau_0 \\ 
\frac{\tau_0}{\beta} + \frac{\tau-\tau_0}{\beta\,\xi} & \tau_0 < \tau \le \tau_0+\Delta \tau  \\
\frac{\tau-\Delta \tau}{\beta} +\frac{\Delta \tau}{\beta \xi}  & \tau > \tau_0+\Delta \tau. 
\end{cases}
\end{eqnarray}

Here, $\tau_0\equiv \tau(t_0)=\ln \left[(1-I_0)/S_b)\right]$, while $\Delta \tau$ has to be determined by integrating $\dot{\tau}=\xi \beta \,I$ during the NPI, which yields the implicit relation \cite{Atias2025}
\begin{equation}
\label{integI}
\int_{\tau_0}^{\tau_0+\Delta \tau} \frac{1}{I(\tau')}\, d\tau'  = \beta \xi \Delta t. 
\end{equation}
Note that $R(\tau)$ in Eq.(\ref{IRdetau}) is continuous and piecewise linear, as expected, while the formula for $I(\tau)$ is just $I=1-S-R$. 

In Ref. \cite{Atias2025}, the authors have also calculated the value of $R_\infty={\rm lim}_{t\to\infty} R(t)$ as
\begin{equation}
\label{Rinfty}
R_\infty=1+\frac{1}{\beta} W_0\left[-\beta e^{-\beta} (1-I_0) e^{(\xi^{-1}-1)\Delta \tau}\right].
\end{equation}
\noindent where $W_0$ is the main branch of Lambert's function. 

Until now, we have presented the mathematical formulations developed in \cite{Atias2025} that are relevant to our studies. In the following section, we turn to our analysis of the critical points and the development of analytical approximations. 

\section{Infection peaks and critical points}

Here we study the critical points and the extrema of $I(t)$, as well as their dependence on the system parameters. Given that $\dot{I}=(dI/d\tau)\dot{\tau}=(dI/d\tau) {\cal B} I$ with ${\cal B} I>0$, the extrema of $I(t)$ can only occur at the values of $\tau$ for which $dI/d\tau=0$ or at the discontinuities found at $\tau=\tau_0$ and $\tau=\tau_0+\Delta \tau$. We first recall that in the absence of NPI, provided that $\beta (1-I_0)>1$, there is a single maximum of $I(t)$ that occurs at the herd immunity threshold $1/\beta=S(t)$. We call this the {\it natural} peak of the epidemics. By considering Eqs.(\ref{IRdetau}) with $\tau_0 \to \infty$ (or either $\xi=1$) it is easy to see that such a maximum is attained for $\tau=\ln(\beta (1-I_0))$ (when $dI/d\tau=0$). Now, for the general case with NPI, taking into account the whole expression for $I(\tau)$ in Eq.(\ref{IRdetau}), we find the following possible critical points of $I(\tau)$, some of which may not exist depending on the parameters.
\begin{itemize}

\item $\tau=\tau_1\equiv \ln(\beta (1-I_0))$. As said, this value of $\tau$ corresponds to the natural peak of the epidemics when there is no NPI. Now, turning to the system with NPI, it is easy to see that $\tau_1$ corresponds to a zero of $dI/d\tau$ if $\tau_1<\tau_0$ or if $\tau_1>\tau_0+\Delta \tau$ but not if $\tau_0<\tau_1<\tau_0+\Delta \tau$. Hence, $\tau_1$ is a critical point that can occur before or after the NPI period, but not within it. We remark that if the parameters are such that $\tau_0<\ln(\beta (1-I_0))<\tau_0+\Delta \tau$, then $\tau_1$ is not a critical point. On the other hand, it can be shown that 
if $\tau_1$ is critical (either before or after the NPI), it always corresponds to a maximum of $I(t)$. In fact, it is easy to see that $d^2 I/d t^2$ is always negative at $t_1\equiv t(\tau_1)$.

\item $\tau=\tau_0=\ln \left[(1-I_0)/S_b)\right]$. This corresponds to the discontinuity of $\dot{I}$ occurring at the beginning of the NPI. It may either correspond to a maximum of $I(t)$ or just to a change of slope, depending on the parameters, but not to a minimum. Specifically, by studying the lateral derivatives of $I(t)$ at $t(\tau_0)$ it can be shown that if $1/(\xi \beta)<S_b$ then $\dot{I}>0$ holds both before and after $\tau_0$. In this case the discontinuity implies a sudden decreasing of the positive 
slope of $I(t)$. Meanwhile, if $1/\beta<S_b<1/(\xi \beta)$, $\tau_0$ corresponds to a maximum of $I$ since $dI/d\tau$ changes from positive to negative. Finally, $S_b<1/\beta$ means that the NPI begins after the occurrence of the natural peak of the epidemics, and hence we have 
$\dot{I}<0$ before and after $t(\tau_0)$. In this latter case, the critical point corresponds to a finite change of the negative slope of $I(t)$.

\item $\tau=\tau_{\xi}\equiv \ln(\xi \beta (1-I_0))$. This is a critical point inside the NPI period as it corresponds to a zero of $dI/d\tau$
in the region $\tau_0<\tau<\tau_{\Delta}$, in contrast to what happens with $\tau_1$. Hence, $\tau_{\xi}\equiv \ln(\xi \beta (1-I_0))$ is critical only if $\tau_0< \tau_\xi <\tau_{\Delta}$. However, as it happens with $\tau_1$, it is easy to show that if $\tau_{\xi}$ is critical, it always corresponds to a maximum. Note that the condition $\tau_0<\tau_\xi$ can be rewritten as $ 1/(\xi \beta)<S_b$,
which is the same condition that ensures that $\dot{I}>0$ holds before and after $\tau_0$. This means that if $\tau_{\xi}$ is critical, then 
$\tau_0$ is not a maximum and $\dot{I}>0$ holds around $t(\tau_0)$. This is also consistent with the fact that, as $\tau_{\xi}<\tau_1$ always holds, $\tau_{\xi}$ can be critical only if there is no previous maximum. 

\item $\tau=\tau_\Delta \equiv \tau_0+ \Delta \tau=\ln \left[(1-I_0)/S_b)\right]+\Delta \tau$. This corresponds to the discontinuity at the end of the NPI. As we indicate below, it can be a minimum of $I(t)$ or a change in slope, but not a maximum. It is easy to show that if $S_b<\exp(\Delta \tau)/\beta$, then $\tau_{\Delta}$ corresponds to a finite slope change with $\dot{I}<0$ on both sides of $\tau_{\Delta}$. Meanwhile, the condition to observe a minimum at $\tau_{\Delta}$ is $\exp(\Delta \tau)/\beta<S_b<\exp(\Delta \tau)/(\beta\xi)$. Finally, if $S_b>\exp(\Delta \tau)/(\beta\xi)$ (equivalent to
$\tau_{\xi}>\tau_{\Delta}$) we have $I(t)>0$ on both sides of $\tau_\Delta$. In the latter case, the NPI is so weak or so early and short that no maximum of $I(t)$ occurs before the end of the NPI (\textit{i.e.} $\tau_0$ corresponds to a change in the positive slope while $\tau_{\xi}>\tau_{\Delta}$ is not critical and the only maximum is at $\tau_1>\tau_{\Delta}$).
\end{itemize}

\begin{figure}
\includegraphics[width=0.9\columnwidth]{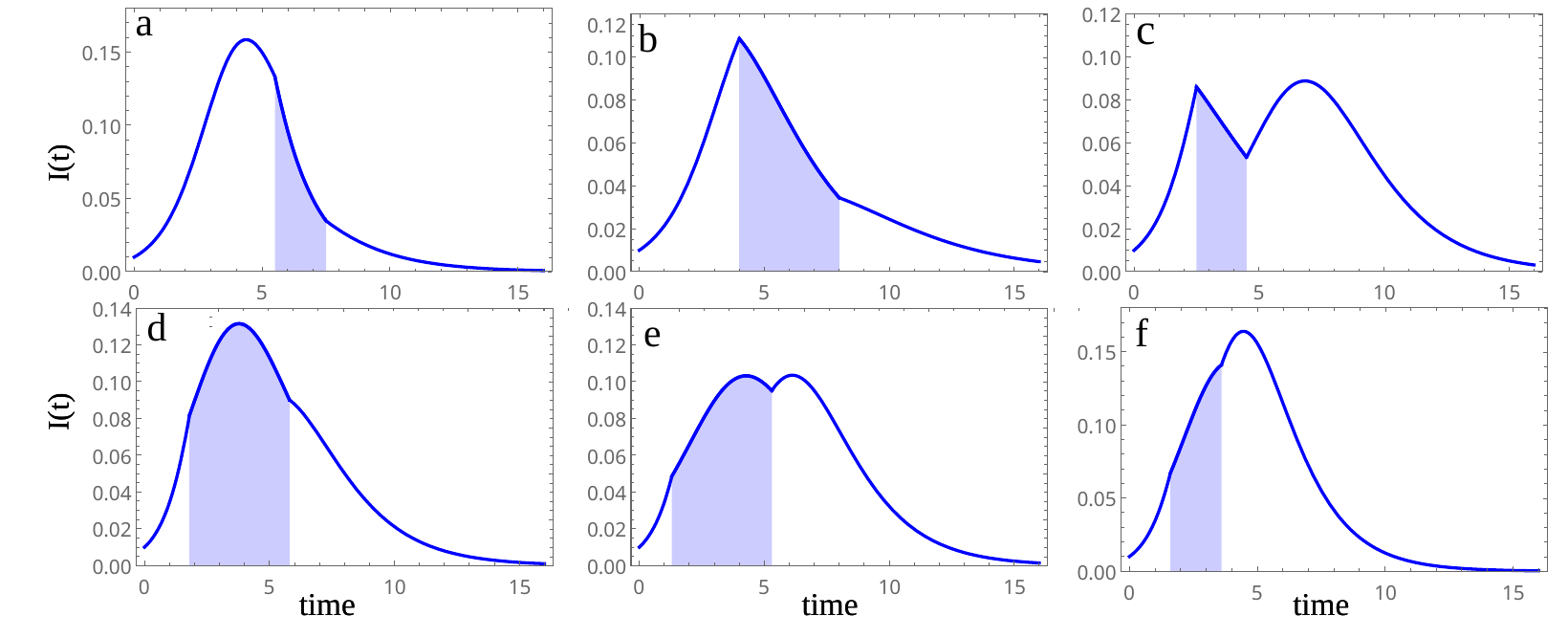}
\caption{\label{figIpfofiles} Typical profiles for $I(t)$ in the different possible scenarios. a) $\tau_{\xi}<\tau_1<\tau_0<\tau_\Delta$. b) $\tau_{\xi}<\tau_0<\tau_1<\tau_{\Delta}$. c) $\tau_\xi<\tau_0<\tau_{\Delta}<\tau_1$. d) 
$\tau_0<\tau_\xi<\tau_1<\tau_{\Delta}$. e) $\tau_0<\tau_\xi<\tau_{\Delta}<\tau_1$. f) $\tau_0<\tau_{\Delta}<\tau_{\xi}<\tau_1$. In all the cases, the time window corresponding to the NPI is shaded. The particular parameters used in the different cases are: $(\beta, \xi, \Delta t, S_b) =  
(2., 0.5, 3, 0.36)$ (a), $(1.8, 0.7, 4, 0.69)$ (b), $(2., 0.5, 2, 0.82)$ (c), $
(2.3, 0.75, 4, 0.85)$ (d), $(2.3, 0.7, 4, 0.92)$ (e), $(2.3, 0.8, 2, 0.88)$ (f). The initial value of the infected fraction is $I_0=0.01$ in all the cases.}
\end{figure}

Interestingly, all the conditions found for each of the values of $\tau_i$ (with $i=0,1,\xi,\Delta$) being critical or not, as well as those determining whether they correspond to maxima, minimum or finite slope changes, can be written in terms of the values of $\tau_i$ themselves. Moreover, such conditions are completely determined by the relative order of the values of $\tau_i$. For instance, the condition $1/(\xi \beta)<S_b$ is just $\tau_0<\tau_1$, the condition $S_b>\exp(\Delta \tau)/(\beta\xi)$ is equivalent to $\tau_{\xi}>\tau_{\Delta}$, etc. In addition, since $\tau_0<\tau_\Delta$ and $\tau_\xi<\tau_1$ always hold, it is easy to show that there are only six possible scenarios, each corresponding to a different allowed order of the values of $\tau_i$. Such six scenarios are illustrated in Fig.~\ref{figIpfofiles}, where we plot $I(t)$ for some selected sets of parameters.
In Fig.~\ref{figIpfofiles}.a we show the case $\tau_{\xi}<\tau_1<\tau_0<\tau_\Delta$ for which the only maximum is at $t(\tau_1)$, before the NPI. In this case, the maximum is just the natural peak of the epidemic, and the NPI is applied later. In panel Fig.~\ref{figIpfofiles}.b we show a typical case with $\tau_{\xi}<\tau_0<\tau_1<\tau_{\Delta}$, for which $\tau_1$ and 
$\tau_{\xi}$ are not critical and the only maximum is at $t(\tau_0)$. Fig.~\ref{figIpfofiles}.c shows the case $\tau_\xi<\tau_0<\tau_{\Delta}<\tau_1$ for which $\tau_0$ and $\tau_1$ are maxima, $\tau_{\Delta}$ is minimum and $\tau_\xi$ is not critical. Panel d shows the case 
$\tau_0<\tau_\xi<\tau_1<\tau_{\Delta}$ for which $\tau_{\xi}$ is the only maximum. The curve in Fig.~\ref{figIpfofiles}.e corresponds to
a case with $\tau_0<\tau_\xi<\tau_{\Delta}<\tau_1$ for which $\tau_\xi$ and $\tau_1$ are maxima separated by a minimum at $\tau_{\Delta}$.
Finally, panel f shows the case $\tau_0<\tau_{\Delta}<\tau_{\xi}<\tau_1$ mentioned at the end of the description of $\tau_{\Delta}$ for which the only maximum is at $\tau_1$, after the NPI. From now on, we refer to these allowed scenarios as $A, B,\dots, F$, according to the panel in which they are portrayed. 

Concerning the maxima of $I(t)$, note that according to the allowed scenarios, if $I(t)$ achieves a single maximum during the evolution of the epidemics, such a maximum can occur before the NPI ($A$), at the beginning of the NPI ($B$), during the NPI ($D$) or after the NPI ($F$). Meanwhile, if there are two peaks, the second one always occurs after the NPI, while the first one may occur during the NPI ($E$) or at the beginning ($C$), but not before. Hence, for example, it is not possible to observe one maximum before the NPI and another after the NPI, nor to find two peaks during the NPI. 

Using the results in Eq.(\ref{IRdetau}) we can calculate the values of $I(t)$ at the critical points. In particular, those that correspond to peaks of $I(t)$. We get
\begin{eqnarray}
\label{Itaui}
I(\tau_1)&=&\begin{cases} \left(\beta-1-\ln[(1-I_0)\beta]\right)/\beta &{\rm for\,\,} \tau_1<\tau_0 \\
\left((\beta-1+\Delta \tau)\xi-\Delta \tau -\xi \ln[(1-I_0)\beta] \right)/(\xi \beta) &{\rm for}\,\, \tau_1>\tau_{\Delta}, \end{cases} \nonumber \\
I(\tau_0)&=&1-S_b-\frac{\ln[(1-I_0)/S_b]}{\beta}, \nonumber \\
I(\tau_{\xi})&=&\frac{\xi \beta -1 - \ln[(1-I_0)\xi \beta]+(1-\xi)\ln[(1-I_0)/S_b]}{\xi \beta} \,\,\,\, {\rm for\,\,} \tau_0<\tau_{\xi}<\tau_{\Delta}, \nonumber \\
I(\tau_{\Delta})&=&\frac{\beta \xi (1-\exp(-\Delta \tau)S_b)-\Delta \tau - \xi \ln[(1-I_0)/S_b]}{\xi \beta}.
\end{eqnarray}

\hspace{\fill} 

Importantly, note that $I(\tau_0)$ and $I(\tau_{\xi})$ are calculated in an exact way in terms of the system parameters $\beta, S_b, I_0$ and $\xi$, and the same occurs with $I(\tau_1)$ if $\tau_1<\tau_0$. In contrast, $I(\tau_{\Delta})$ depends on $\Delta \tau$ (as well as $I(\tau_1)$ do if $\tau_1>\tau_{\Delta}$), and there is not an explicit expression for $\Delta \tau$ as a function $\Delta t$ and the other system parameters. We only have the implicit relation given in Eq.(\ref{integI}). This problem also affects the calculation of $R_\infty$ (see Eq.(\ref{Rinfty})). In the following section we provide an approximate explicit expression for $\Delta \tau$ that helps in our analysis of the infection peaks and $R_\infty$ as functions of the system parameters. 

\section{Analytical approximation for $\Delta \tau$}

The major drawback of the implicit equation for calculating $\Delta \tau$ given in Eq.(\ref{integI}) is the fact that there is no known solution for the integral involved. To deal with this, in Ref. \cite{Atias2025} the authors performed a Taylor expansion on $\Delta \tau$ that enabled them to find relevant results for the minimization of 
$R_{\infty}$ as a function of $S_b$ at small $\Delta t$. Here, we propose a different way of treating the problem. 

First, by plugging the expression for $I(\tau)$ given in Eq.(\ref{IRdetau}) into Eq.(\ref{integI}) and changing the integration variable to $u=\tau'-\tau_0$, we get
\begin{equation}
\label{integu}
\int_0^{\Delta \tau}
\frac{1}{1-\frac{\tau_0}{\beta}-(1-I_0) e^{-\tau_0}\, e^{-u}-\frac{u}{\xi \beta}} \, du = \beta \xi \Delta t. 
\end{equation}
Here, it should be noted that the main difficulty for solving the integral is the fact
that the denominator of the integrand includes the linear term $(-u/(\beta\xi))$ summed to a term proportional to $e^{-u}$. Now, recalling that $\sinh(u)\simeq u$ for $u\ll 1$ (since $\sinh(u)=u+u^3/6+o(u^5)$), we make the ansatz of replacing the linear term $-u/(\beta\xi)$ in Eq.(\ref{integu}) by $(-\sinh(u)/(\beta\xi))$, expecting that this may lead to a good approximation for small enough $\Delta \tau$. With this substitution, the integral can be solved since the denominator of the integrand becomes a linear combination of exponential functions plus a constant. In this way, we get
\begin{equation}
\label{equalitysinh}
\frac{
2 e^{\tau_0/2} \, \xi \left[
\arctan\left( 
\frac{e^{\tau_0/2} (1 - \xi \beta + \xi \tau_0)}
{\sqrt{2 \xi \beta s_0 - e^{\tau_0} \left(1 + \xi^2 (\beta - \tau_0)^2 \right)}}
\right)
-
\arctan\left(
\frac{e^{\tau_0/2} (e^{\Delta \tau} - \xi \beta + \xi \tau_0)}
{\sqrt{2 \xi \beta s_0 - e^{\tau_0} \left(1 + \xi^2 (\beta - \tau_0)^2 \right)}}
\right)
\right] \beta
}{
\sqrt{2 \xi \beta s_0 - e^{\tau_0} \left(1 + \xi^2 (\beta - \tau_0)^2 \right)}
} \simeq \beta \xi \Delta t,
\end{equation}
where $s_0=(1-I_0)$, and $\tau_0=\ln \left[(1-I_0)/S_b)\right]$ as previously indicated. 
It is worth stressing that the left hand side of Eq. (\ref{equalitysinh}) is the exact result of the integral in Eq.(\ref{integu}) when considering the replacement mentioned before. Now, Eq. (\ref{equalitysinh}) can be easily solved to obtain 
an explicit (approximate) expression for $\Delta \tau$ as a function of $\beta, \xi, S_b, \Delta t$ and $I_0$. This is 
\begin{equation}
\label{explDtau}
\Delta \tau\simeq \ln \left[A_1 \tan\left[\tan^{-1}\left(\frac{-\beta \xi +\xi \ln \left(\frac{1-I_0}{S_b}\right)+1}{A_1}\right) -\frac{1}{2} \Delta t A_1\right]+\xi \left(\beta -\ln
  \left(\frac{1-I_0}{S_b}\right)\right)\right],
\end{equation}
where
\begin{equation}
\label{A1}
A_1=\sqrt{\xi ^2 \ln \left(\frac{1-I_0}{S_b}\right) \left(2 \beta -\ln \left(\frac{1-I_0}{S_b}\right)\right)+\beta \xi (2
  S_b-\beta \xi )-1}.
\end{equation}

Here we have expressed $\tau_0$ as a function of the system parameters to get the explicit formula, but note that both expressions in Eqs.(\ref{explDtau}) and (\ref{A1}) can be simplified if written in terms of $\tau_0$. 

In the following sections, we use the relation given in Eq.(\ref{explDtau}) to analyze the dependence of the extrema of $I(t)$ and $R_{\infty}$ on the system parameters. For this, we just plug the approximation for $\Delta \tau$ into Eqs. (\ref{Itaui}) and (\ref{Rinfty}).

Although the approximation made to solve the integral in Eq.~(\ref{integu}) seems rather uncontrolled, note that if the value found for $\Delta \tau$ in a given example is small enough, by consistence, we expect that the integral should be well approximated with the replacement $u\to \sinh(u)$ because the values of $u$ in the integration interval will be small. Thus, the procedure should be sound and the results for $\Delta \tau$ should be accurate. In contrast, if the value found for $\Delta \tau$ in Eq. (\ref{explDtau}) is not small, it is probably not a good approximation. As we show in the examples in the next section, 
$\Delta \tau \lesssim 1/3$ leads to good approximations for the different quantities analyzed. In contrast, larger values of $\Delta \tau$ usually produce non-negligible errors. Notably, as we will see, small values of $\Delta \tau$ do not necessarily imply small values of $\Delta t$, and the approximation given in Eq.~(\ref{explDtau}) is rather good for a wide and relevant region of parameters, although not for any parameter set. 

\begin{figure}
\includegraphics[width=0.95\columnwidth]{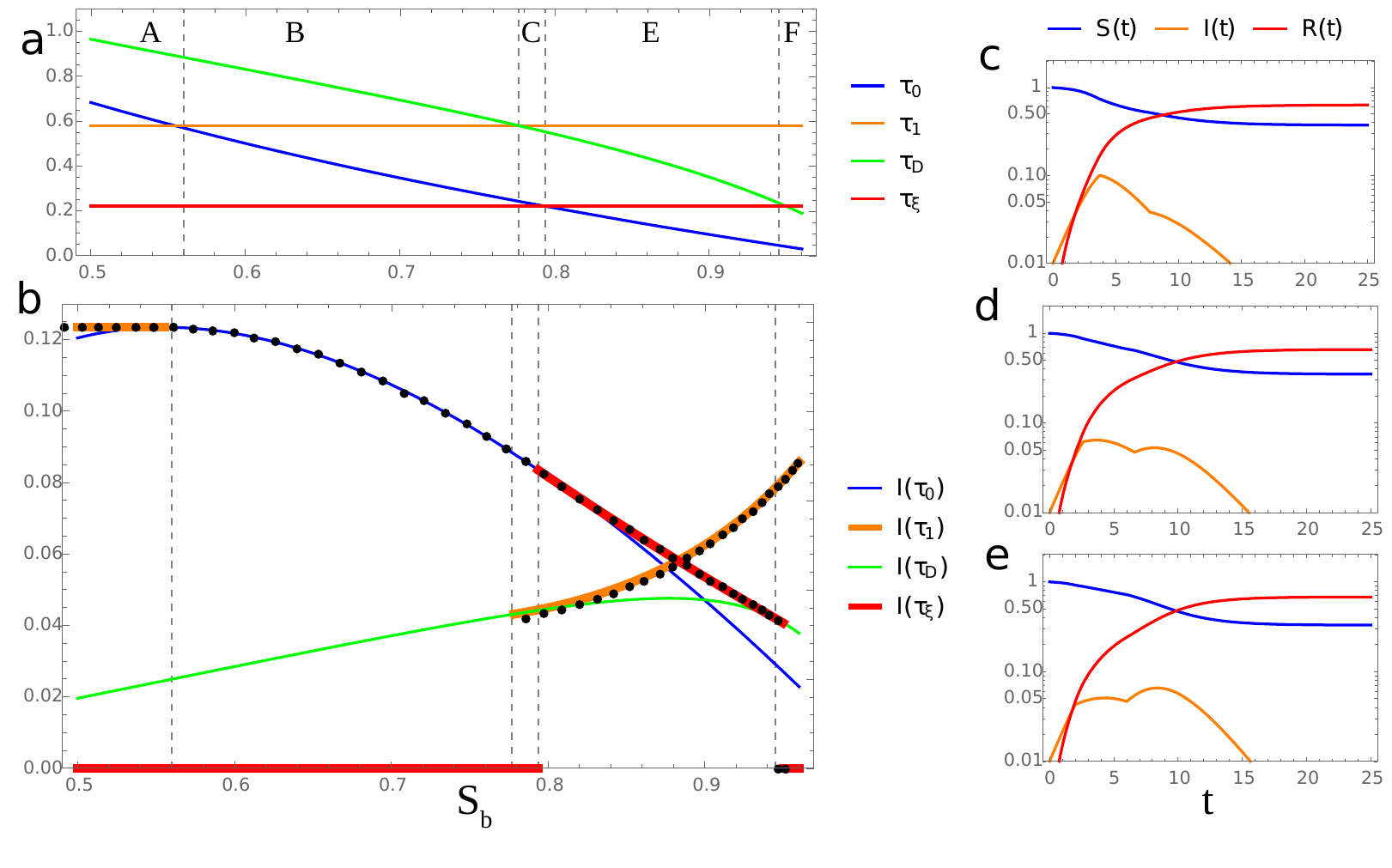}
\caption{\label{fig2IdeSb} a) Values of $\tau_i$ as functions of $S_b$ for $\beta=1.8, \xi=0.7, \Delta t=4, I_0=0.01$. b) Values of $I(t)$ at the
critical points as function of $S_b$ for the same systems shown in panel (a). The solid colored curves indicate the analytical solutions, while the black dots are the numerical solutions for the local maxima found for the corresponding set of parameters. The vertical dashed lines in panels (a) and (b) indicate the changes of the observed scenario. From left to right the corresponding scenarios are $A, B, C, E$ and $F$, as indicated at the top of panel (a). Panels (c), (d) and (e) show the evolution of $I(t),S(t)$ and $R(t)$ for the systems analyzed in panel (b) considering $S_b=0.73$ (scenario $B$), $S_b=0.86$ (scenario $E$) and $S_b=0.91$ (scenario $E$), respectively. For convenience, the results for $I(\tau_1)$ and $I(\tau_\xi)$ in (b) are set as equal to zero in the zones where $\tau_1$ and $\tau_\xi$ are not critical, respectively.}
\end{figure}

\section{Results for temporary NPIs in the standard SIR model}

Here we study the dependence of the maxima of $I(t)$ and $R_\infty$ on the system parameters, comparing analytical approximations with numerical solutions. In particular, we check to what extent the approximation made for calculating $\Delta \tau$ is valid.

First, we analyze the influence of the parameter $S_b$. Fig.~\ref{fig2IdeSb}.a shows the values of $\tau_i$ ($i=0, 1, \xi, \Delta$) as functions of $S_b$ for fixed values of $\beta, \Delta t, \xi$ and $I_0$. Here, $\Delta \tau$ is calculated analytically with the approximation presented in the previous section. We see that, as $S_b$ varies from $0.55$ to $0.95$, the values of the critical points change in such a way that the solution scans the scenarios $A,B,C,E$ and $F$, skipping scenario $D$. Each change of scenario is determined by a crossover of two critical points and
is indicated with a vertical dashed line. In Fig. \ref{fig2IdeSb}.b we plot the values of $I(\tau_i)$ for all the critical points in the same range of $S_b$ considered in Fig.~\ref{fig2IdeSb}.a. The solid lines correspond to calculations using Eqs. (\ref{Itaui}) while the black dots indicate the maximal values of 
$I(t)$ found in numerical solutions of the dynamical equations. We see that the analytical calculations correctly predict the numerical results for the maximal values of $I(t)$. In particular, note that the results for $I(\tau_1)$ for the scenarios $C, E$ and $F$ involve the use of the approximation for $\Delta \tau$. We stress the fact that numerical results for $I(\tau_i)$ are shown only for the maximal values of $I$. This means neither for the minimum that occurs at 
$\tau_\Delta$ in scenarios $C$ and $E$, nor for the cases in which the critical points $\tau_0$ or $\tau_\Delta$ involve no extrema but only slope changes.
For convenience, the results for $I(\tau_1)$ and $I(\tau_\xi)$ are set as equal to zero in the zones where $\tau_1$ and $\tau_\xi$ are not critical, respectively. Recall that $I(\tau_0)$ is a maximum only in scenarios $B$ and $C$ while in the others it corresponds
to a slope change. Meanwhile, $I(\tau_1)$ is a maximum in scenarios $A,C,E$ and $F$ and is not critical in $B$. On the other hand, $I(\tau_\xi)$ is a maximum for scenario $E$ (and for $D$ which does not appear for the parameters studied here), and not critical elsewhere. Finally, $I(\tau_\Delta)$ is minimum for scenarios $C$ and $D$ and corresponds to a slope change elsewhere. 

In Figs.~\ref{fig2IdeSb}.c,~\ref{fig2IdeSb}.d and~\ref{fig2IdeSb}.e, we show the evolution of the dynamical variables for three particular values
of $S_b$ with the rest of the parameters as in Fig. \ref{fig2IdeSb}.b. The curves in Figs. \ref{fig2IdeSb}.c correspond to a typical example in scenario $B$ with the only maximum at $t_0$. Meanwhile, in Figs. \ref{fig2IdeSb}.d and Figs. \ref{fig2IdeSb}.e we show two examples of epidemic profiles 
in scenario $E$ with different relative heights of the two peaks. That is, $I(\tau_\xi)>I(\tau_1)$ in panel (d) and $I(\tau_\xi)<I(\tau_1)$ in panel (e). Note that as $S_b$ increases within the range of scenario $D$, the first infection peak (located at $\tau_\xi$) decreases, while the second one (at $\tau_1$) increases. The minimal absolute maximum of $I(t)$ is obtained when $I(\tau_\xi)=I(\tau_1)$, at $S_b\simeq 0.88$

As the parameter $S_b$ is directly related to the time at which the NPI begins, the dependence of the results on $S_b$ may be relevant from the point of view of the management of epidemics \cite{Atias2025}. For example, according to the results in Fig. \ref{fig2IdeSb}.b, if the decision makers managing the epidemics have estimated the values of $\beta$ and $\xi$ of the order of those used in our calculations, and they plan to impose a quarantine with the considered value of $\Delta t$, then, starting the quarantine at $S_b\sim 0.88$ would be optimal in minimizing the height of the infection peaks, although understanding that there would be two peaks. 

\begin{figure}
\includegraphics[width=0.5\columnwidth]{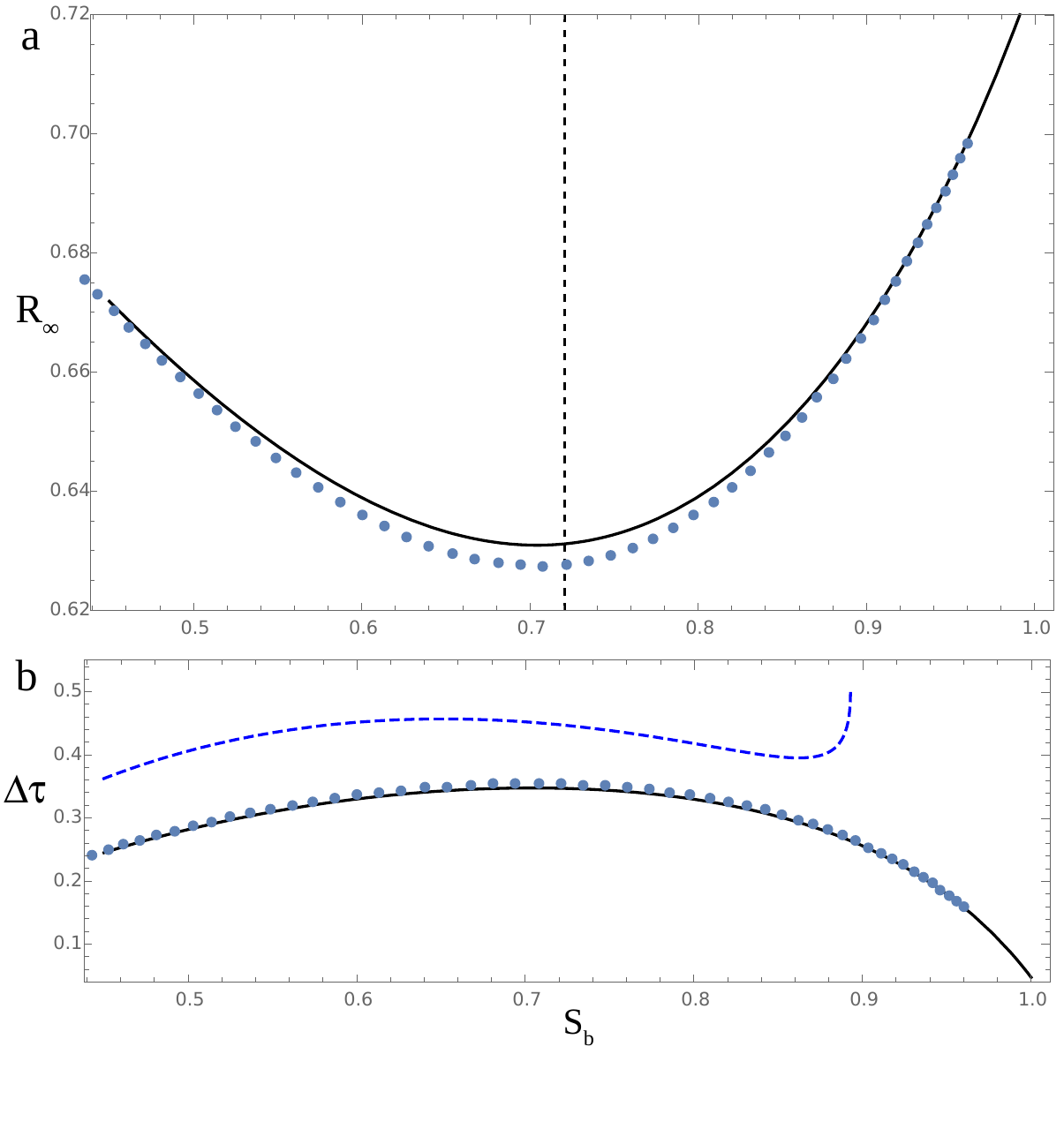}
\caption{\label{figRinf1} a) $R_\infty$ as a function of $S_b$ for $\beta=1.8, \xi=0.7, \Delta t=4, I_0=0.01$. The solid line corresponds to the
analytical approximation while the dots correspond to numerical solutions of the dynamical equations. The dashed vertical line indicates the approximate 
position of the minima according to the result given in \cite{Atias2025}. b) $\Delta \tau$ as a function of $S_b$ for the same parameters as in panel (a). The black solid line corresponds to the analytical approximation 
taken from Eq.(\ref{equalitysinh}), the dots correspond to the numerical solution, the blue dashed line is for an approximation considering a Taylor expansion (see text).}
\end{figure}

Now we turn to the analysis of the behavior of $R_\infty$. In Fig. \ref{figRinf1}.a we show the dependence of $R_\infty$ on $S_b$ for the same systems and parameters analyzed in Fig. \ref{fig2IdeSb}. The analytical approximation is computed using the (exact) formula given in Eq.(\ref{Rinfty}) with the approximate result for $\Delta \tau$ taken from Eq.(\ref{equalitysinh}). We see that the agreement between the analytical approximation and the numerical results is quite good, but there are some small differences for intermediate values of $S_b$. The origin of such differences can be understood by observing Fig. \ref{figRinf1}.b. Let us first focus on the solid black curve and the circles, which plot the results for $\Delta \tau$ as a function of $S_b$ corresponding to the analytical approximations and the numerical solutions, respectively. Note that, for intermediate values of $S_b$, $\Delta \tau$ increases up to values of order $\sim 0.3$. In such zones, the difference between $\sinh(u)$ and $u$ begins to influence the calculations so that tiny divergences between the approximation for $\Delta \tau$ and the numerical results are observed. We argue that this also affects the results for $R_\infty$. It is worth noting that very small differences between analytical and numerical calculations are also observed in the results for $I(\tau_1)$ in Fig.~\ref{fig2IdeSb}.b for the region close $S_b\sim 0.8$ that can be attributed to the same cause. 

Regarding the shape of the curve $R_\infty$ vs. $S_b$, it is important to note that the minimum of $R_\infty$ is attained for 
$S_b\simeq 0.7$. This is much smaller than the value $S_b\sim 0.88$ needed to minimize the maximal infection peak. Moreover, according to the results in Fig.~\ref{fig2IdeSb}.b, for $S_b\simeq 0.7$ there is a quite large infection maximum ($I(\tau_1)\simeq 0.105$) which almost doubles the peaks found
at $S_b\sim 0.88$. Hence, as expected, planning an NPI for minimizing $R_\infty$ may require quite different actions that aiming at reducing the infection peak. On the other hand, the value $R_\infty\simeq$ found when optimizing $I_{max}$ is only about $5\%$ higher than that found when 
optimizing $R_\infty\,$ ($0.66$ vs $0.63$). This suggests that in this case it may be better to focus on reducing the peaks than on reducing $R_{\infty}$, however, this may change with the parameters $R, \xi$ and $\Delta t$. 

Finally, in Fig.~\ref{figRinf1} we also show results obtained by using an approximations of the kind considered in \cite{Atias2025}. First, in 
Fig.~\ref{figRinf1}.a, the vertical dashed line indicates the result obtained for the minimum of $R_{\infty}$ as a function of $S_b$ 
according to the formula given in \cite{Atias2025}. That is,
\begin{equation}
\label{SoptAtias}
S_{opt}\simeq \frac{1}{\beta}+\frac{(\beta-1-\ln(\beta))\,\xi \Delta t}{2 \beta}.
\end{equation}
As we see, the approximation is quite good for the parameters considered here. The actual optimal value found is $S_{opt}\simeq 0.703$, while Eq.(\ref{SoptAtias}) yields $S_{opt}\simeq 0.72$. On the other hand, in Fig.~\ref{figRinf1}.b, the dashed blue line shows the results for $\Delta \tau$ 
as a function of $S_b$ obtained by using a Taylor expansion of Eq. (\ref{integI}) for small $\Delta \tau$ in the way it is indicated in \cite{Atias2025} (see comment in \cite{comment1}). It is apparent that the solid line for the approximation in Eq. (\ref{equalitysinh})
provides much better results than the Taylor approximation. Moreover, note that the solution from the Taylor expansion ceases to be valid for $S_b\gtrsim 0.89$. This is because it becomes complex (\textit{i.e.} with a non-vanishing imaginary part), as it results from a second order equation whose discriminant becomes negative. 

\begin{figure}
\includegraphics[width=0.95\columnwidth]{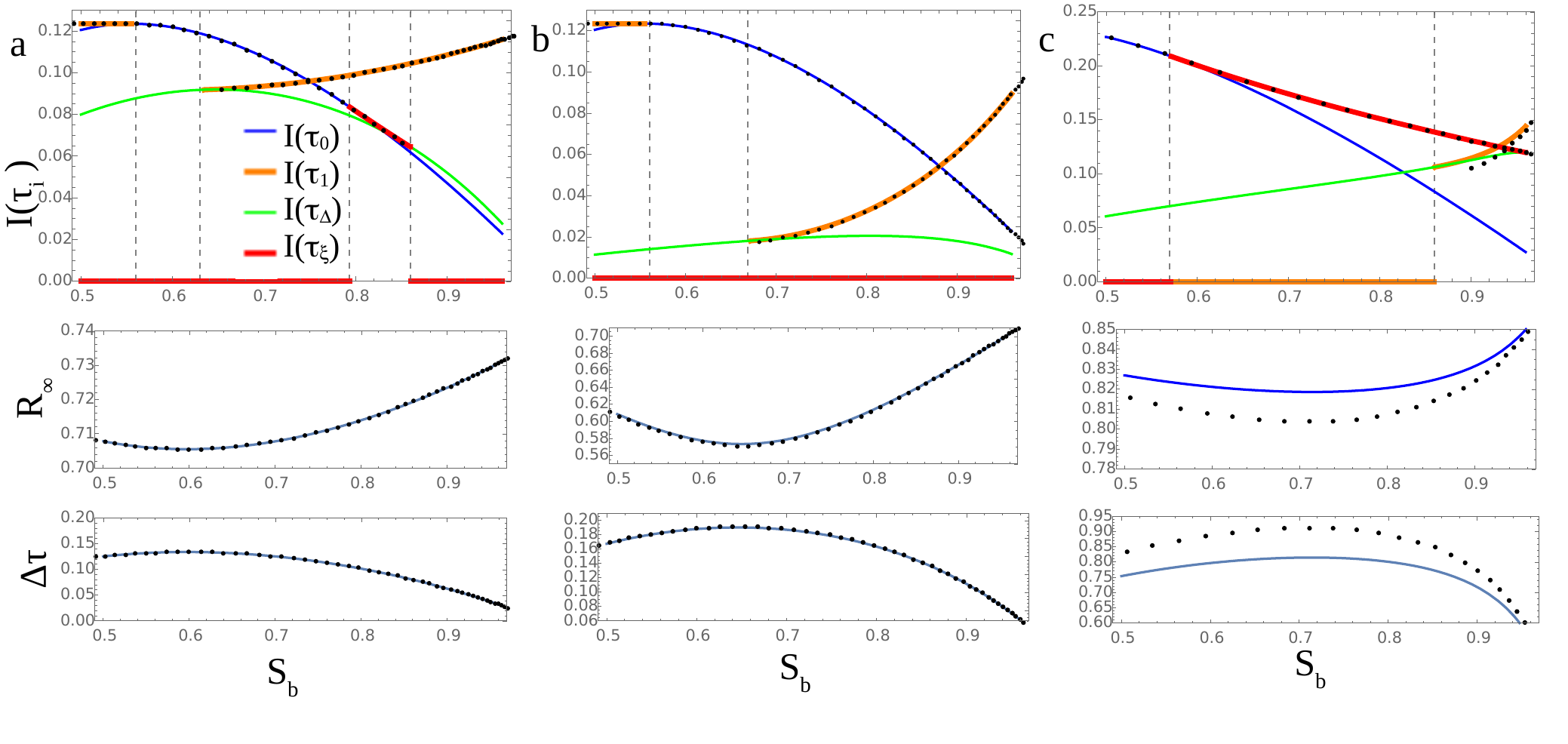}
\caption{\label{fig4} Each panel a, b, and c, shows the infection level at critical points (top), $R_\infty$ (center) and $\Delta \tau$ (bottom) as functions of $S_b$. In all the cases, the solid curves correspond to analytical approximations while the dots correspond to numerical solutions. The top panels include numerical results only for the local maxima, not for the rest of the critical points. 
The parameters are the same as in Fig. \ref{fig2IdeSb} excepting for $\Delta t=1$ (a), $\xi=0.5$ (b), and $\beta=2.5$ (c). The vertical doted lines in the top panels indicate the changes in the observed scenarios. In panel (a) we have scenarios $A,\, B,\, C,\, E,\, F$ from left to right, in panel (b) scenarios $A,\, B,\, C$, and in (c) scenarios $B,\,D,\,E$.}
\end{figure}

In Fig.~\ref{fig4} we show results for $I(t)$ at the critical points as functions of $S_b$, and also for $R_\infty$ and $\Delta \tau$ considering
three sets of parameters different from the one used in Figs. \ref{fig2IdeSb} and \ref{figRinf1}. We see that as $S_b$ varies in the range $0.5 - 0.95$, different sequences of scenarios are scanned depending on the other parameters. While in the case analyzed in Fig.~\ref{fig4}.a we found the
same sequence of scenarios as in Fig. \ref{fig2IdeSb} (namely, $A, B, C, E, F$), the systems studied in Fig.~\ref{fig4}.b scans only the scenarios $A, B, C$,
and the systems in Fig.~\ref{fig4}. explores the scenarios $B, D, E$. In the three cases, the value of $S_b$ that minimizes $R_\infty$ is considerably lower than that of the one that minimizes the maximal infection peak. In the first two cases (Figs. \ref{fig4}.a and \ref{fig4}.b) the values of $\Delta \tau$ shown in the lower panels are relatively small in the whole range of $S_b$ analyzed ($\Delta \tau < 0.2$). Thus, the analytical approximations 
for $\Delta \tau$, $R_\infty$ and the infection peaks agree very well with the numerical results. In contrast, in the example in Fig. \ref{fig4}.c, 
the values of $\Delta \tau$ are relatively large $\Delta \tau>0.5$ and, hence, the approximations for $\Delta \tau$ and $R_\infty$ exhibit relevant
differences from the numerical results. This also leads to non negligible errors in the prediction of the values of $I(\tau_1)$ for $S_b\sim 0.9$. 
Importantly, note that the inexactness of the prediction of $\Delta \tau$ does not affect the values of $I(\tau_\xi)$ and $I(\tau_0)$ that are the
relevant values in most of the scan range of $S_b$. 

\begin{figure}
\includegraphics[width=0.95\columnwidth]{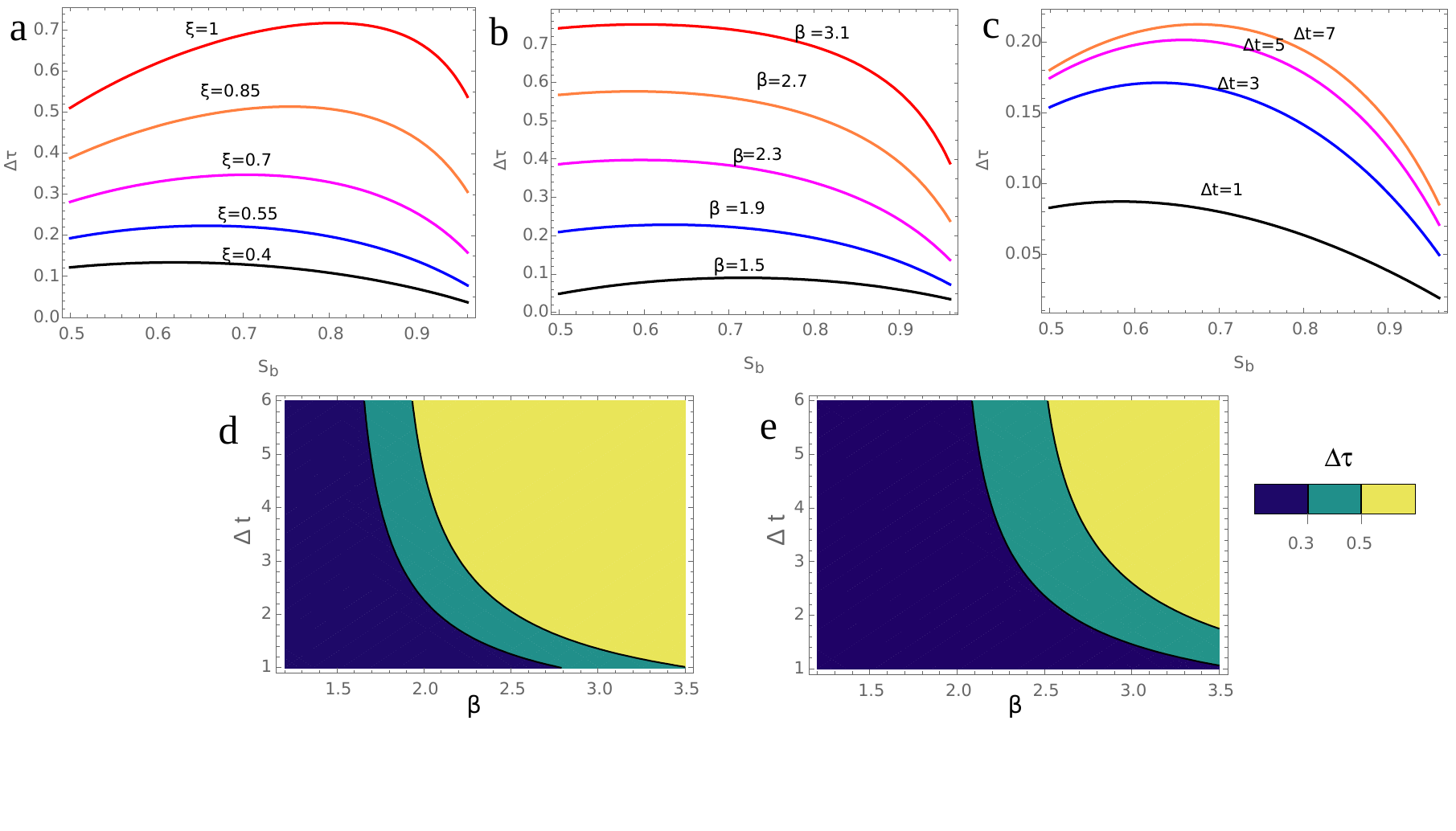}
\caption{\label{figdeltatau} Analytical approximation for $\Delta \tau$ in Eq.(\ref{explDtau}). a) $\Delta \tau$ as a function of $S_b$ for various values of $\xi$, considering $\beta=1.8,\, \Delta t=4$. b) $\Delta \tau$ as a function of $S_b$ for various values of $\beta$,
considering $\xi=0.5,\, \Delta t=4$. c) $\Delta \tau$ as a function of $S_b$ for various values of $\Delta t$,
considering $\beta=1.8,\, \xi=0.5$. d) Contour plot of $\Delta \tau$ as a function of $\beta$ and $\Delta t$ considering $\xi=0.7$ and $S_b=0.8$. 
The contours indicate the levels $\Delta \tau=0.3$ and $\Delta \tau=0.5$. e) Ibid (d) for $\xi=0.5$. All the calculations consider $I_0=0.01$.}
\end{figure}

The examples shown as well as others not shown indicate that values of $\Delta \tau\lesssim 0.3$ are well predicted by the analytical approximation and lead to good results for the infection peaks. In contrast $\Delta \tau \gtrsim 0.5$ produces non negligible errors. To understand in which parameter regions the approximation for $\Delta \tau$ performs well, Fig. \ref{figdeltatau} analyzes the dependence of $\Delta \tau$ on the various parameters. Fig. \ref{figdeltatau}.a shows the approximation for $\Delta \tau$ given in Eq.(\ref{explDtau}) as a function of $S_b$ for different values of $\xi$. The results indicate that $\Delta \tau$ increases with $\xi$. Hence, the smaller $\xi$, the better the approximation. Similarly, the results in Figs. \ref{figdeltatau}.b and \ref{figdeltatau}.c indicate that the approximation improves with decreasing $\beta$ and decreasing $\Delta t$. In Figs. \ref{figdeltatau}.d and \ref{figdeltatau}.e we show contour plots of the analytical result for $\Delta \tau$ as function of $\beta$ and $\Delta t$ for two different values of $\xi$ and fixed $S_b=0.8$. In the regions where $\Delta \tau < 0.3$, the approximation is expected to perform well.

\FloatBarrier
\section{Temporary NPIs in Network systems: individual and environmental measures vs quarantines.}

We now extend the analysis to network models, with two main goals. On the one hand, we address up to what point the general results for varying $S_b$ found using the standard SIR model still hold for NPIs in network systems; on the other hand, we compare the effects of the two types of NPIs mentioned in the introduction: reducing the mean contagion rate, and taking lockdown measures in order to reduce the mean contact rate in the population. These are two essentially different intervention types that cannot be distinguished using the standard SIR models, but can be directly implemented in complex network based models.

\subsection{Network model and assumptions}

Throughout this Section, we use the degree-based mean field model (DBMF) proposed by Moreno et.al.~\cite{Moreno1} with a slight modification presented in \cite{Barthelemy}. The model represents the population as a complex network with degree distribution $P(k)$, where the degree $k$ of an individual represents the amount of daily contacts that can lead to a contagion from an infectious to a susceptible person. 

This model considers a sub-division of the compartments for each degree. Namely, $S_k(t)$, $I_k(t)$ and $R_k(t)$ stand for the fraction of indivduals with degree $k$ that are in the S, I and R compartments, respectively, at time $t$. This leads to $S_k+I_k+R_k=1$ at any given time and for any given degree $k$. The evolution of the infection is given by the following equations:

\begin{eqnarray}
\frac{d S_k}{dt} &=& -\beta k S_k\Theta \nonumber \\
\frac{d I_k}{dt} &=& \beta k S_k\Theta - I_k \label{network_model}\\
\frac{d R_k}{dt} &=& I_k \nonumber
\end{eqnarray}

\noindent where the time has been rescaled as in Eq.~(\ref{SIReqs}), $\beta$ is the ratio between the contagion and recovery rates, and $\displaystyle \Theta(t) = \sum_k \frac{P(k)\,(k-1)}{\langle k\rangle} I_k(t)$ is the probability that any given link points to an infected individual, regardless of their degree, in uncorrelated networks. See more details in~\cite{Moreno1}. Numerical solutions of  Eqs.~(\ref{network_model}) were performed using the custom Python code available in Ref. \cite{DataAv}.  Throughout this section, we will show results for the total fraction of individuals in each compartment. For example, the total fraction of infected individuals can be computed as
\begin{equation}
I(t)=\sum_{k=1}^K P(k) \: I_k(t)
\end{equation}
\noindent while $S(t)$ and $R(t)$ can be obtained in analogous ways.

The basic reproduction number of this network model is $\mathcal R_0=\beta\frac{\langle k^2\rangle-\langle k\rangle}{\langle k\rangle}$ \cite{Andersson,Rozan2}.
In our studies, we focus on Poisson networks, which have degree distribution $P(k)=\exp{(-\lambda)}\, \lambda^k/k!$, where the parameter $\lambda$ coincides with the mean number of links of the agents (i.e. $\lambda=\langle k \rangle$). The reproduction number yields $\cal R_0=\beta\,\lambda$ \cite{Rozan2}.
Note that the model has an additional parameter ($\lambda$) compared to the standard SIR. In our analysis, we sometimes consider $\mathcal R_0$ instead of $\lambda$ as the relevant parameter.

As mentioned above, the model allows for the implementation of NPIs in two different ways. First, during the NPI period, the infection rate can be reduced in a factor $0<\xi_\beta<1$ by considering the change $\beta \to \xi_\beta \beta$, as done in the standard SIR model. This is a preferable assumption to model individual and/or environmental interventions that reduce transmission without affecting the structure of social contacts. Relevant analytic and numerical results regarding this case were provided in~\cite{Atias2025}. Second, we can consider a change in the distribution of contacts $P(k)\to \tilde{P}(k)$ during the NPI period, as done in \cite{Rozan,Svoboda}. This is more suitable for modeling lockdown or quarantine measures. When limiting to Poisson networks, the appropriate change is $\lambda \to \xi_\lambda \lambda$ with $0<\xi_\lambda<1$, so that the mean number of contacts is reduced during the NPI. It is important to mention that the procedure for implementing the change of the degree distribution in the middle of the numerical integration is non-trivial and is explained in detail in \cite{Rozan}. 
Finally, we can consider NPIs that involve both actions together, \textit{i.e}. modifying $\beta$ and $\lambda$ at the same time.

In principle, we could analyze NPIs with any values of $\xi_\beta$ and $\xi_\lambda$ and compare their influence on the dynamics of the epidemic. For example, for a given set of parameters, comparing the results of applying an NPI with $\xi_\beta=0.9$ and $\xi_\lambda=0.9$ with those of an NPI with $\xi_\beta=1$ and $\xi_\lambda=0.5$, would mean comparing the influence of an NPI that combines weak individual protective measures and weak lockdown measures with that of a hard quarantine with no individual measures. For simplicity, in this work we limit the analysis to the following three cases:
\begin{itemize}
\item Case 1 --- Only individual measures: $(\xi_\beta,\xi_\lambda)=(\xi,1)$.
\item Case 2 --- Only lockdown measures: $(\xi_\beta,\xi_\lambda)=(1,\xi)$. 
\item Case 3 --- Combination of measures: $(\xi_\beta,\,\xi_\lambda)=(\sqrt{\xi},\,\sqrt{\xi})$,
\end{itemize}
where the same value of $\xi$ ($0<\xi<1$) is considered in the three cases. Note that the comparison between Cases 1 and 2 is relevant since this corresponds to comparing the effects of lowering the contagion rate with those of decreasing the average number of contacts by the same factor. In fact, both cases lead (formally) to the same change
$\mathcal R_0 \to \xi \mathcal R_0$ (since $\mathcal R_0=\beta \lambda$). This means that both would produce the same effects on the initial outbreak if the NPI was applied at the beginning of the epidemic. From another point of view, at any time, they would both affect in similar ways the initial contagions induced by a single infected agent within a neighborhood of purely susceptible individuals. 
Clearly, the combination considered in Case 3 leads to the same change in the product $\beta\cdot\lambda$.

\subsection{Results for network systems.}
With the considerations stated above, we analyze the numerical solutions of the system. To begin, we find numerically the infection peaks and analyze them as functions of $S_b$. For this, we first search for the local maximum of $I(t)$ in the time range $0 \leq t \leq t_0$, including the possibility of finding it exactly at $t_0$ (\textit{i.e.}, at the beginning of the NPI). We call this maximum $I_a$. In principle, our algorithm considers the possibility of finding more than one maximum in this region, but we always found only one, as expected. Then, we do the same scan process for the time range $t_0\leq t \leq t_\Delta$, and call the found maximum $I_b$. Next, we consider the range $t_\Delta \leq t$ and call the maximum there $I_c$. 

In addition, we register the values $I_0$ and $I_\Delta$ that represent the fraction of infected individuals at the start and at the end of the NPI, respectively. Note that in some cases there could be coincidences: we may get $I_a=I_0$, $I_b=I_0$, $I_c=I_\Delta$ and so on. 

\begin{figure}[ht]\centering
\begin{subfigure}{.49\textwidth} \centering
\includegraphics[width=\linewidth]{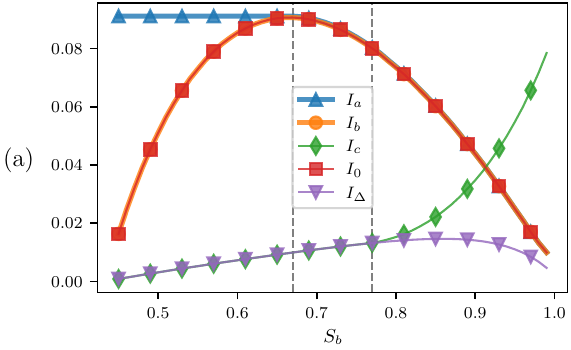}
\end{subfigure}
\begin{subfigure}{.49\textwidth} \centering
\includegraphics[width=\linewidth]{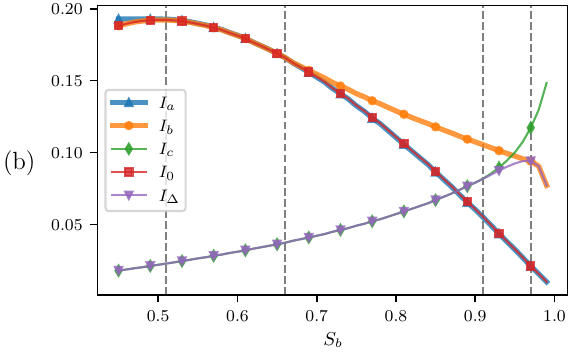}
\end{subfigure}
\vspace{0.05cm}
\begin{subfigure}{\textwidth} \centering
\includegraphics[width=\linewidth]{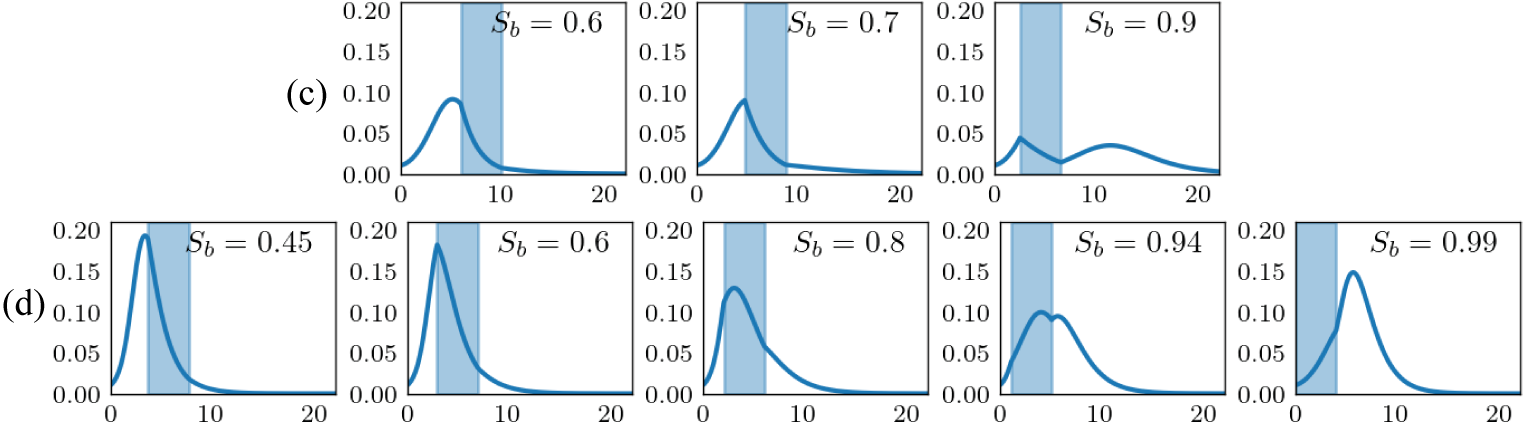}
\end{subfigure}
\caption{Values of $I(t)$ at critical times (see text for details) as functions of $S_b$. a) Parameter values $\lambda = 3.6, \beta = 0.5$ 
(yielding $\mathcal R_0=1.8$), $\Delta t=4$ and $(\xi_\beta,\xi_\lambda) = (0.5,1)$. The vertical dashed lines indicate the transition between scenarios $A, B, C$ from left to right. b) Parameter values $\beta = 0.5, \lambda =5$ (resulting in $\mathcal R_0=2.5$), $\Delta t=4$ and $(\xi_\beta,\xi_\lambda) = (0.7,1)$. The vertical lines separate the scenarios observed, which are $A, B, D, E, F$ from left to right. c) Profiles of $I(t)$ for three selected values of $S_b$ for the system studied in panel (a). Each value of $S_b$ corresponds to a different scenario. d) The same as panel (c) for five values of $S_b$ taken from the system studied in panel (b). Each one corresponds to a different scenario.}
\label{network_peak_beta}
\end{figure}

Fig.~\ref{network_peak_beta} shows results for the critical values of $I(t)$ mentioned as functions of $S_b$, when the NPI consists of reducing $\beta$ only (Case 1 described above).

 In Fig.~\ref{network_peak_beta}.a, we first see that for small values of $S_b$ ($S_b<0.67$) the NPI starts after the natural peak of infection. A particular example is plotted in the left panel of Fig.~\ref{network_peak_beta}.c. Clearly, this corresponds to scenario $A$ in Fig \ref{figIpfofiles}. For moderate values of $S_b$ (between $0.67$ and $0.77$), the NPI starts before the epidemic could reach its natural peak. The relatively strong reduction in the infection rate considered here ($\xi_\beta=0.5$) makes the fraction of infected individuals decrease monotonously during the NPI, and therefore we have $I_b=I_0$ and the infection profiles correspond to the scenario $B$. A particular example is shown in the central panel of Fig.~\ref{network_peak_beta}.c. Finally, for $S_b>0.77$, there is a second peak of infections after the NPI, since $I_c$ is higher than $I_\Delta$. This corresponds to scenario $C$, as shown in the example in the right panel of Fig.~\ref{network_peak_beta}.c. 
On the other hand, in Fig.~\ref{network_peak_beta}.b, in contrast to the behavior observed in Fig.~\ref{network_peak_beta}.a, there are values of $S_b$ for which there is an infection peak during the NPI, since $I_b>I_0$. This occurs because both $\mathcal R_0$ and $\xi$ are larger than in the previous case. Thus, at the beginning of the NPI, infected individuals are still able to spread the disease faster than they recover, therefore increasing $I(t)$. For even higher $S_b$, the fraction of susceptible individuals is still high enough so that the infection curve could continue to increase at the end of the NPI, and so $I_\Delta = I_b$. The solutions scan the scenarios $A, B, D, E$ and $F$
as $S_b$ increases.

In summary, for Case 1 NPIs we observe very similar behaviors as those observed in Fig.~\ref{fig4} for the standard SIR model. In particular, the same types of scenario and sequences of scenarios arise for varying $S_b$. This coincidence is not necessarily trivial, since the models are not the same. In fact, we cannot completely discard the possibility that the network model produces other types of scenario for other parameter sets.

\begin{figure}[ht]\centering
\begin{subfigure}{.49\textwidth} \centering
\includegraphics[width=\linewidth]{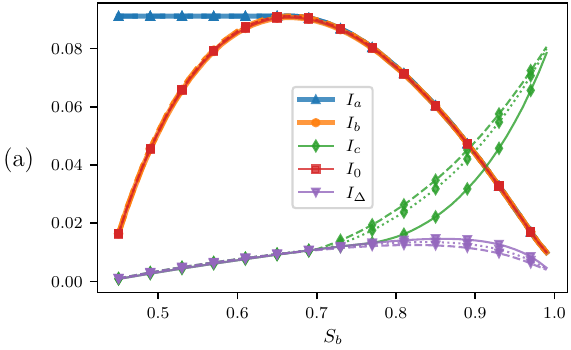}
\end{subfigure}
\begin{subfigure}{.49\textwidth} \centering
\includegraphics[width=\linewidth]{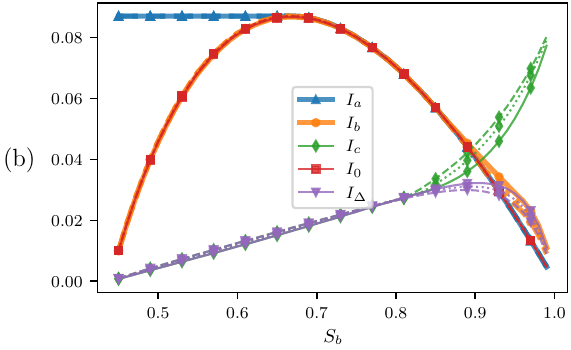}
\end{subfigure}
\begin{subfigure}{.49\textwidth} \centering
\includegraphics[width=\linewidth]{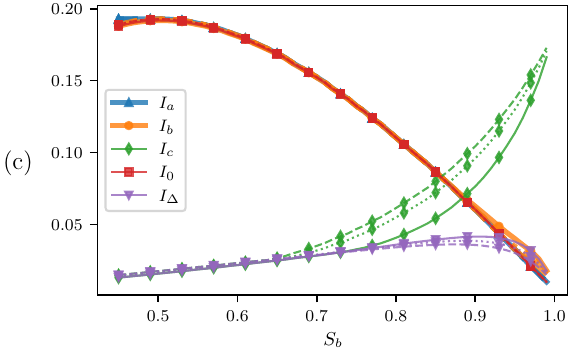}
\end{subfigure}
\begin{subfigure}{.49\textwidth} \centering
\includegraphics[width=\linewidth]{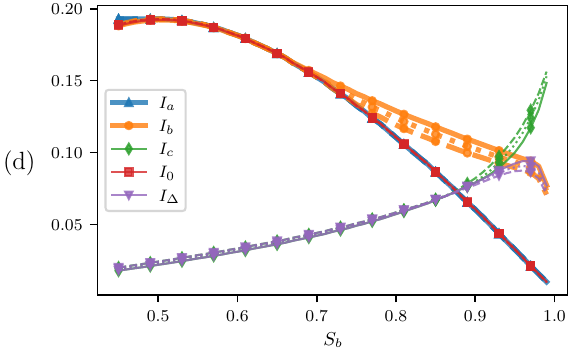}
\end{subfigure}
\caption{Critical values of $I(t)$ as functions of $S_b$ in network systems considering the three cases of NPIs studied. In each panel, solid lines corresponds to the Case 1 ($\beta\to \xi \beta$), dashed lines correspond to Case 2 ($\lambda\to \xi \lambda$), while dotted lines correspond to Case 3 ($(\beta,\lambda)\to(\sqrt{\xi}\beta,\sqrt{\xi}\lambda)$). The parameters are $I_0=0.01, \beta=0.5, \Delta t=4$ considering $\lambda = 3.6, \xi = 0.5$ (a), $\lambda = 3.6, \xi = 0.7$ (b), $\lambda = 5, \xi = 0.5$(c), $\lambda = 5, \xi = 0.7$ (d).}
\label{network_peaks_cases}
\end{figure}

We now proceed to compare the different types of NPIs considering the three cases mentioned above. That is, reducing $\beta$, reducing $\lambda$, and reducing both at the same time. 
In Fig.~\ref{network_peaks_cases}, we analyze the influence of each type of NPI on the infection peaks considering four different sets of parameters and scanning the relevant ranges of $S_b$. Fig.~\ref{network_peaks_cases}.a reproduces the results for Case 1 shown in Fig. \ref{network_peak_beta}.a and also includes the results for Case 2 and Case 3 for the same system parameters. First, it should be stressed that only the critical points that occur after the beginning of the NPI may change with the type of NPI applied, \textit{i.e.} neither $I_a$ nor $I_0$ can vary, but $I_b, I_\Delta$ and $I_c$ do. 

\begin{figure}[ht]\centering
\begin{subfigure}{.49\textwidth} \centering
\includegraphics[width=\linewidth]{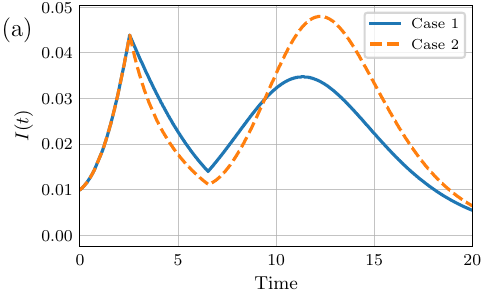}
\end{subfigure}
\begin{subfigure}{.49\textwidth} \centering
\includegraphics[width=\linewidth]{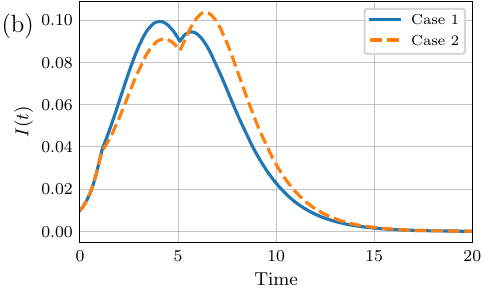}
\end{subfigure}
\caption{Profiles of $I(t)$ for particular parameter set selected from the systems studied in Fig.~\ref{network_peaks_cases}. (a): $\lambda=3.6,\xi = 0.5, S_b=0.90$. (b): $\lambda=5,\xi = 0.7, S_b=0.94$. The rest of the parameters are the same as in Fig.~\ref{network_peaks_cases}. }
\label{compara_I}
\end{figure}

In fact, relevant changes are observed. The value of $I_c$ (\textit{i.e.} the peak after the NPI) for Cases 2 and Case 3 is considerably higher than for Case 1. For example, for $S_b\sim0.8$ the $I_c$ peak of Case 2 almost doubles that of Case 1. As expected, the results for Case 3 are between those for Cases 1 and 2. But we see that they are closer to those for Case 2. In addition, in Fig.~\ref{network_peaks_cases}.a it can be seen that the sequence of scenarios found as $S_b$ increases is the same for the three cases. This is $A$, $B$, $C$ from left to right. However, we have not indicated the transitions with vertical lines as in Fig.~\ref{network_peak_beta} because they occur at different values of $S_b$ for each Case considered. Inclusion of all lines could cause confusion. In particular, note that the transition from $B$ to $C$ for Cases 2 and 3 (\textit{i.e.} the separation between $I_c$ and $I_\Delta$) occurs at lower values of $S_b$ than for Case 1. In relation to this, we see the important fact that the minimization of the maximal infection peak for Case 2 requires a lower value of $S_b$ than for Case 1. Note that the minimum peak occurs at the value of $S_b$ for which $I_c=I_b$, which is slightly lower than $0.9$ for Case 2 and slightly higher than $0.9$ for Case 1. 
This means that Case 2 NPI should be imposed later than Case 1 NPI if one aims at minimizing the absolute maximum of $I(t)$. 
Finally, we note that there are small but non negligible changes in the values of $I_\Delta$ for the different cases, which in scenario $C$ corresponds to the minimum between the two peaks. We see that, for cases 2 and 3 the minimum is deeper than for case 1. Hence, Case 1 produces a flatter curve with a lower second peak and a less deep valley between the two maxima than the other cases. This is shown in Fig.~\ref{compara_I}.a for a particular example. Now, going back to Fig.~\ref{network_peaks_cases}, for the case studied in Fig.~\ref{network_peaks_cases}.b, the parameters are such that the sequence of scenarios is $A$, $B$, $C$, $E$, $F$ for the three cases. Again, the maximum after the NPI period ($I_c$) is smaller for Case 1. This is observed for scenarios $C$, $E$ and $F$. Meanwhile, in scenario $E$ (occurring in a narrow window at $S_b\gtrsim.9$) the peak inside the NPI period ($I_b$) is similar for the three cases. The value of $I_\Delta$ in scenarios $C$ and $E$ (\textit{i.e.} the minimum between the two infection peaks), is always lower for case 2, as in Fig.~\ref{network_peaks_cases}.a. 
In addition, the value of $S_b$ that minimizes the absolute maximum shifts to the left as in the previous case. The results in Fig.~\ref{network_peaks_cases}.c are similar to those in Fig.~\ref{network_peaks_cases}.b, with no novelties despite the difference in the parameters. In Fig.~\ref{network_peaks_cases}.d we consider the same parameters as in ~\ref{network_peak_beta}.b, so that the curves for Case 1 are the same, and here we add those for Cases 2 and 3. 
The scenarios found are $A$, $B$, $D$, $E$, $F$ as indicated for ~\ref{network_peak_beta}.b. Regarding the peak after the NPI ($I_c$), the results are similar to those in Figs.~\ref{network_peaks_cases}.a-c. This means that the peak is lower for Case 1 in both scenarios $D$ and $E$. In contrast, the peak $I_b$ presents the opposite behavior, \textit{i.e.} it is lower for Case 2 and higher for Case 1. Note that $I_b$ occurs during the NPI period and corresponds to the only peak in scenario $D$ ($S_b\lesssim 0.9$), while it is the first of the two peaks in scenario $E$ ($0.9\lesssim S_b \lesssim 0.97$). The value of $S_b$ needed to minimize the absolute maximum of $I(t)$ is again smaller for Case 2, although this is difficult to see without zooming the figure. Fig.~\ref{compara_I}.b shows the profiles found for Cases 1 and 2 for a particular value of $S_b$ in scenario E.

In summary, Case 1 causes a lower second peak than Case 2, while Case 2 produces a lower first peak if such a peak appears inside the NPI period. Moreover, if the parameters are such that the $I(t)$ profile has only one peak that occurs during the NPI (scenario D), such a peak is smaller for Case 2. The minimization of the absolute maximum of $I(t)$ occurs at lower values of $S_b$ for Case 2. The results for Case 3 are always between those for Case 1 and Case 2. 

\begin{figure}[ht]\centering
\begin{subfigure}{.49\textwidth} \centering
\includegraphics[width=\linewidth]{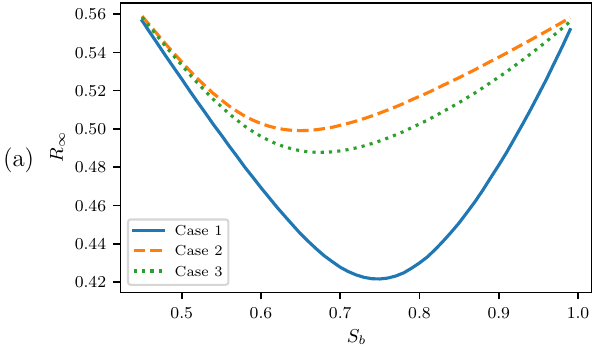}
\end{subfigure}
\begin{subfigure}{.49\textwidth} \centering
\includegraphics[width=\linewidth]{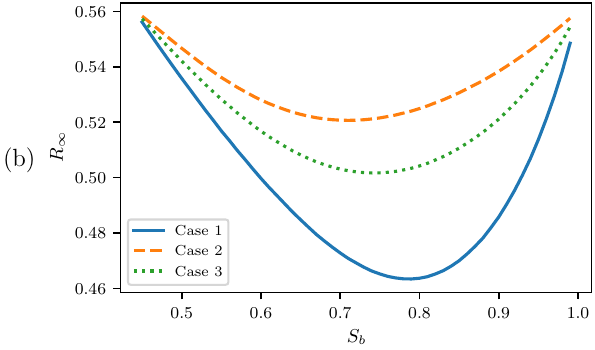}
\end{subfigure}
\begin{subfigure}{.49\textwidth} \centering
\includegraphics[width=\linewidth]{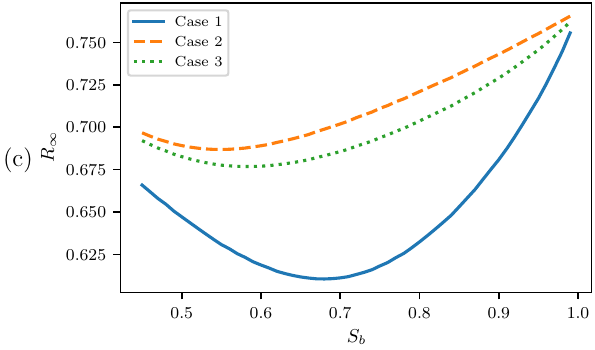}
\end{subfigure}
\begin{subfigure}{.49\textwidth} \centering
\includegraphics[width=\linewidth]{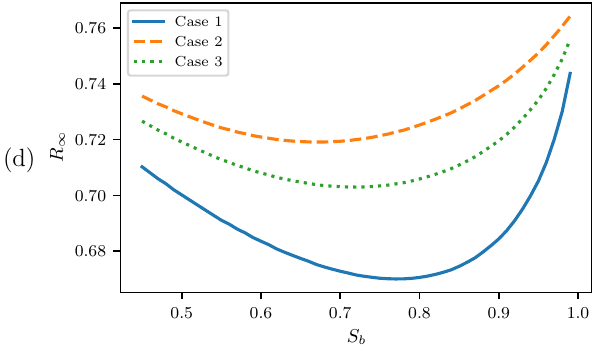}
\end{subfigure}
\caption{Outbreak size $R_\infty$ as a function of $S_b$ for the four sets of parameters explored in Fig.~\ref{network_peaks_cases}. The parameters $\lambda$ and $\xi$ in each panel are the same as in the corresponding panel in Fig.~\ref{network_peaks_cases}. }
\label{network_Rinf}
\end{figure} 

Now we turn to the analysis of the values of $R_\infty$ found for the different types of NPIs. In Fig.~\ref{network_Rinf} we show the results for 
$R_\infty$ as functions of $S_b$ for the same systems and cases studied in Fig.~\ref{network_peaks_cases}. Three main things have to be noted. First, as happens for the ordinary SIR model without network, the value of $S_b$ that minimizes $R_\infty$ is always significantly smaller than that minimizing the absolute maximum of $I(t)$. In all the examples studied, a value $S_b\simeq 0.9$ is needed to minimize the peak while the optimal $S_b$ for minimizing $R_\infty$ is always lower than $0.8$. This means that the NPI has to be applied earlier if one aims to minimize the peak instead of $R_\infty$. Second, the values found of $R_\infty$ are considerably lower using the strategy of Case 1. Third, the optimal value of $S_b$ concerning the minimization of $R_\infty$ or the minimization of the peak is considerably lower for Case 2 than for Case 1. This means that, in general, the optimal time $t_0$ to begin the NPI for Case 2 is larger than for Case 1.

\section{Conclusions}

In this work, we have explored the influence of different types of NPIs within the framework of a standard SIR model, and also by considering a degree-based mean-field network model. Our studies follow and complement those provided in \cite{Atias2025}. 

Concerning the standard SIR model with NPIs, we have presented analytical results for the critical points and the maximal values of $I(t)$ as functions of the duration of the NPI ($\Delta t$), the infection rate ($\beta$), the strength of the NPI ($\xi$) and the fraction of susceptible agents at which the NPI is implemented ($S_b$). 

Among other relevant features, the results reveal the existence of six different allowed scenarios for the evolution of the epidemic in what concerns the number of infection peaks before, during, and after the NPI.
The analytical results are exact for the critical points found before and during the NPI, while we provide approximations for the critical points that occur after the NPI. We also indicate how the reliability of the approximations depends on the parameters. 

Our results show that for a fixed duration of the NPI, the value of $S_b$ that minimizes $R_\infty$ is found to be lower than that needed to minimize the infection peak, for all the cases analyzed. Regarding epidemic management, this means that the NPI should be applied earlier if the objective is to minimize the peak, compared to the case in which the relevant quantity to reduce is $R_\infty$.
However, for various parameters sets analyzed, minimizing the peaks leads to values of $R_\infty$ that are not much higher than the optimal one. In contrast, minimizing $R_\infty$ frequently leads to rather high peaks. From this point of view, the minimization of the peaks could seem to be a more efficient strategy. Nevertheless, this may may depend on the parameters. 

The analytical treatment and approximations provided, especially concerning the solution for the integral in Eq.(\ref{integu}), may be of interest for future work in related topics, since the same integral appears, for instance, in the studies in \cite{Atias2025} and \cite{julicher2020}. 

The network model allowed us to distinguish between two main types of NPIs. On one hand, those that involve only individual and/or environmental measures that decrease the transmission rate (Case 1), and on the other hand, those that change the contact structure of the population, reducing the number of social contacts (Case 2). Our analysis shows that, when considering the same duration of the NPI and the same assumptions for their effects on the basic reproductive number, Case 1 leads to considerably lower values of $R_\infty$ and lower heights  of the second peak than Case 2, while Case 2 produces lower values of the first peak. In general, the optimal time to begin the NPI for Case 2 is greater than for Case 1, both for peak minimization or final outbreak minimization.

In a given epidemiological situation, the preferable interventions will always depend on the details of the disease, the particularities of the affected populations, and also on the available resources. No single strategy
is universally optimal. 
Our studies complement those of \cite{Atias2025} in providing a framework for evaluating NPIs based on simple models. Together with other works in the field, they can integrate the toolbox of policy makers when timely decisions are required.

\section{Acknowledgments}
The authors acknowledge financial support from CONICET and CNEA. All public Institutions from Argentina.

\end{document}